# Universal quantum correlation close to quantum critical phenomena

Meng Qin[1,2]*, Zhong-Zhou Ren[1,3,4]*, and Xin Zhang[1]

[1]Department of Physics and Key Laboratory of Modern Acoustic, Nanjing University, Nanjing, 210093, China.

[2]College of Sciences, PLA University of Science and Technology, Nanjing, 211101, China.

[3]Center of Theoretical Nuclear Physics, National Laboratory of Heavy-Ion Accelerator, Lanzhou, 730000, China.

[4]Kavli Institute for Theoretical Physics China, Beijing, 100190, China.

*email: qrainm@gmail.com

*email: zren@nju.edu.cn

**ABSTRACT**

We study the ground state quantum correlation of Ising model in a transverse field (ITF) by implementing the quantum renormalization group (QRG) theory. It is shown that various quantum correlation measures and the Clauser-Horne-Shimony-Holt inequality will highlight the critical point related with quantum phase transitions, and demonstrate nonanalytic phenomena and scaling behavior when the size of the systems becomes large. Our results also indicate a universal behavior of the critical exponent of ITF under QRG theory that the critical exponent of different measures is identical, even when the quantities vary from entanglement measures to quantum correlation measures. This means that the two kinds of quantum correlation criterion including the entanglement-separability paradigm and the information-theoretic paradigm have some connections between them. These remarkable behaviors may have important implications on condensed matter physics because the critical exponent directly associates with the correlation length exponent.

Quantum phase transitions (QPTs) signify that the ground state of the many-body system dramatically changes by varying a physical parameter—such as pressure or magnetic field. The one-dimensional Ising model in a transverse field (ITF) [1-6] can be used to explain the phenomena of ferromagnetic, ferroelectric, and order-disorder transformations. It has been obtained comprehensive study as the simplest exactly solvable model to demonstrate QPTs.

Traditionally, the way of studying the phase transitions is the mean field theory. But researchers found that the mean field theory results are not in agreement with the experiments because the mean field theory ignores the effect of fluctuation. It is the quantum fluctuation that induces QPTs. One of the most important progresses happened in 1970 when Wilson introduced the concept of renormalization in the quantum field theory to quantum statistical physics[7]. He used the renormalization theory to investigate the Ising model and derived the universal law of the second order phase transitions through the scale theory and the computation of critical exponent. The result is the most revolutionary breakthrough to find the nature of QPTs[8].

Besides the direct investigation on the relations between entanglement and QPTs in different systems[9-11], combining the quantum renormalization group (QRG) method and the quantum entanglement theory to study quantum critical phenomena also has attracted great attention[12-18]. Many interesting and meaningful results have been got by studying the low dimensional spin system. M. Kargarian[12, 13] have found that the derivative of the concurrence between two blocks each containing half of the system size diverges at the critical point. The behavior of the entanglement near the critical point is directly connected to the quantum critical properties. The divergent behavior of the first derivative of the concurrence was accompanied by the scaling behavior in the vicinity of the critical point[14].

But things have changed dramatically as some new developments took place that the quantum entanglement cannot be viewed as the whole quantum correlation and the only useful resource in quantum information processing[19]. One of the important concepts came up in 2001 when Henderson *et al.* and Ollivier *et al.* have concluded that entanglement does not account for all nonclassical correlations and even those separable states contain correlations that can be demonstrated by quantum discord[19-21]. Inspired by the meaningful results about quantum discord, many similar quantum correlation measures based on information-theoretic have been proposed, such as measurement-induced disturbance, geometric discord, measurement-induced nonlocality, quantum deficit and so on[19, 22-25]. The investigations on these methods also have obtained much attention[26-37]. Among them one of the topics deserves more attention that is the relations between these quantum correlation measures and QPTs. Some of the questions still remain to be solved: are there some general and special properties when we use these quantum correlation measures to study the QPTs if the QRG theory is adopted? How does the scaling behavior or the critical phenomena change as we take different measures? If the answer can give some universal result, this will be very important and meaningful to know the relations between QPTs and quantum information theory because QPTs can be used to recover the qubit in quantum information processing. This paper is our attempt to solve these questions.

**Results**

**Renormalization of the Hamiltonian.** The Hamiltonian of ITF on a periodic chain of $N$ qubits can be written as[12]

$$H = -J \sum_{i=1}^{N} \left( \sigma_i^z \sigma_{i+1}^z + g \sigma_i^x \right), \qquad (1)$$

where $J$ is the exchange interaction, $\sigma_i^\tau$ ($\tau = x, z$) are Pauli operators at site $i$, and $g$ is the transverse field strength. In order to implement QRG, one needs to divide the Hamiltonian into two-site blocks. The Hamiltonian then can be resolved into the block Hamiltonian $H^B = -J \sum_{I=1}^{N/2} \left( \sigma_{1,I}^z \sigma_{2,I}^z + g \sigma_{1,I}^x \right)$ and interacting Hamiltonian $H^{BB} = -J \sum_{I=1}^{N/2} \left( \sigma_{2,I}^z \sigma_{1,I+1}^z + g \sigma_{2,I}^x \right)$, here $\sigma_{j,I}^\alpha$ are the Pauli matrices at site $j$ of the block labelled by $I$.

The two lowest eigenstates of the corresponding $I$th block Hamiltonian $-J\left( \sigma_{1,I}^z \sigma_{2,I}^z + g \sigma_{1,I}^x \right)$ is given by

$$\begin{aligned} |\psi_0\rangle &= \alpha |00\rangle + \beta |11\rangle, \\ |\psi_1\rangle &= \alpha |01\rangle + \beta |10\rangle, \end{aligned} \qquad (2)$$

where $\alpha = s/\sqrt{s^2+1}$, $\beta = 1/\sqrt{s^2+1}$, and $s = \sqrt{g^2+1} + g$. Now we can establish the relations between the original Hamiltonian and the renormalized one that is $H^{eff} = P_0 \left( H^B + H^{BB} \right) P_0$. The projection operator $P_0 = |\psi_0\rangle_{II}\langle \Uparrow| + |\psi_1\rangle_{II}\langle \Downarrow|$ can be constructed by using the two lowest eigenstates $|\psi_0\rangle$ and $|\psi_1\rangle$, where $|\Downarrow\rangle_I$ and $|\Uparrow\rangle_I$ are renamed states of each block to represent the effective site degrees of freedom[12]. The effective Hamiltonian of the renormalized chain is again an ITF model which is similar to the original Hamiltonian $H$

$$H_{eff} = -J' \sum_{i=1}^{N/2} \left( \sigma_i^z \sigma_{i+1}^z + g' \sigma_i^x \right), \qquad (3)$$

where the renormalized couplings are

$$J' = J \frac{2\left( \sqrt{g^2+1} + g \right)}{1 + \left( \sqrt{g^2+1} + g \right)^2}, \quad g' = g^2. \qquad (4)$$

Accordingly, the density matrix of the ground state is given by

$$\rho = |\psi_0\rangle\langle\psi_0| = \begin{bmatrix} \beta^2 & 0 & 0 & \alpha\beta \\ 0 & 0 & 0 & 0 \\ 0 & 0 & 0 & 0 \\ \alpha\beta & 0 & 0 & \alpha^2 \end{bmatrix}. \qquad (5)$$

In the following, after briefly introducing the definitions of different measures, we will investigate the characteristic of them.

**Negativity.** Firstly, we use the negativity to calculate the entanglement in the model. According to the Peres-Horodecki criterion, a non-entangled state has necessarily a positive partial transpose (PPT)[38]. The Peres-Horodecki criterion gives a qualitative way to judge whether the state is entangled or not. Negativity, firstly introduced by Vidal and Werner, is defined by[39]

$$Ne(\rho_{AB}) = \frac{||\rho^{T_A}||_1 - 1}{2}, \quad (6)$$

where $||\rho^{T_A}||_1$ denotes the trace norm of the partial transpose $\rho^{T_A}$,

$$||\rho^{T_A}||_1 = \mathrm{tr}\sqrt{(\rho^{T_A})^\dagger \rho^{T_A}}. \quad (7)$$

It is easy to compute the negativity of eq. (5)

$$Ne(\rho) = \frac{s}{s^2+1} = \frac{1}{2\sqrt{g^2+1}}. \quad (8)$$

The properties of entanglement $Ne$, the first derivative of $Ne$, and the scaling behavior of $|dNe/dg|_{\min}$ have been plotted in Fig. 1. It is shown that the negativity develops two different saturated values with the increasing of the size of the system in Fig. 1(a). The saturated value of negativity is zero corresponding to the paramagnetic phase when the magnetic field parameter is in $1 < g \leq 2.5$, while the saturated value of negativity is 0.5 corresponding to the long-ranged ordered Ising phase for $0 \leq g < 1$. Therefore, the two phases are separated by critical point $g_c = 1$. The nonanalytic feature of the first derivative of negativity at the critical point is given in Fig. 1(b). The system exhibits singular property as the number of QRG iterations increases. In order to give a more detailed analysis, the values of the minimum $|dNe/dg|_{\min}$ as a function of the system size are depicted in Fig. 1(c) after enough times of iteration. It can be seen that the $\ln|dNe/dg|_{\min}$ shows a linear characteristic with $\ln(N)$ and $g_{\min}$ gradually approaches to the critical point $g_c$. The relations for this kinds of conditions are $|dNe/dg|_{\min} \sim N$. So the critical exponent $\theta$ is 1. Ref [12] have demonstrated that the exponent $\theta$ is directly related to the correlation length exponent $\nu$ close to critical point. So we can easily compute the correlation length through QRG theory.

**Quantum discord and measurement-induced disturbance.** Quantum discord $(QD)$[21, 29] is defined by the following expression

$$QD(\rho_{AB}) = I(\rho_{AB}) - CC(\rho_{AB}). \quad (9)$$

For an arbitrary bipartite state $\rho_{AB}$, the total correlations are expressed by quantum mutual information

$$I(\rho_{AB}) = \sum_{i=A,B} S(\rho_i) - S(\rho_{AB}), \qquad (10)$$

where the mutual information measures the total correlation, including both classical and quantum, for a bipartite state $\rho_{AB}$. Here $S(\rho) = -tr(\rho \log_2 \rho)$ denotes the von Neumann entropy, with $\rho_A$ and $\rho_B$ being the reduced density matrix of $\rho_{AB}$ obtained by tracing out $A$ and $B$, respectively. The classical correlation $CC(\rho_{AB})$ is defined as

$$CC(\rho_{AB}) = \max_{\{\Pi_k^B\}} \left[ S(\rho_A) - S(\rho_{A|B}\{\Pi_k^B\}) \right] = S(\rho_A) - \min_{\{\Pi_k^B\}} S(\rho_{A|B}\{\Pi_k^B\}), \qquad (11)$$

where $S(\rho_{A|B}\{\Pi_k^B\}) = \sum_k p_k S(\rho_k)$ is quantum conditional entropy. The maximum is achieved from a complete set of projective measurements $(\Pi_k^B = |B_k\rangle\langle B_k|, k=1,2)$ on subsystem $B$ locally.

Measurement-induced disturbance (*MID*) is defined as the difference of two quantum mutual information respectively of a given state $\rho_{AB}$ shared by two parties (*A* and *B*) and the corresponding post-measurement state $\Pi(\rho_{AB})$ [25, 29]

$$MID(\rho_{AB}) = I(\rho_{AB}) - I(\Pi(\rho_{AB})), \qquad (12)$$

where the mutual information is the same as defined in eq. (10). $I(\Pi(\rho))$ quantifies the classical correlation in $\rho_{AB}$, with $\Pi(\rho_{AB}) = \sum_{i,j} (\Pi_i^A \otimes \Pi_i^B) \rho_{AB} (\Pi_i^A \otimes \Pi_i^B)$, where the measurement is induced by the spectral resolutions of the reduced states $\rho_A = \sum_i p_i^A \Pi_i^A$ and $\rho_B = \sum_i p_i^B \Pi_i^B$.

After some standard algebra, we can get *QD* and *MID* as

$$QD(\rho) = h(p_1) + h\left(\left(\beta^2 - \alpha^2\right)^2\right) - \left(1 + \beta^2 + \alpha^2\right) h(p_2)/2. \qquad (13)$$

$$MID(\rho) = (\alpha^2 + \beta^2)\log_2(\alpha^2 + \beta^2) - \alpha^2 \log_2(\alpha^2) - \beta^2 \log_2(\beta^2). \qquad (14)$$

Here $p_1 = x^2 + \max(t_1^2, t_2^2)$, $p_2 = \left((2\beta^2 - 2\alpha^2)^2 + 16\alpha^2\beta^2\right)/(1 + \beta^2 + \alpha^2)$, $x = \beta^2 - \alpha^2$, $t_1 = 2\alpha\beta$, $t_2 = -2\alpha\beta$, $t_3 = \beta^2 + \alpha^2$. The function $h(z)$ can be expressed as

$$h(z) := -\frac{1+\sqrt{z}}{2}\log_2\frac{1+\sqrt{z}}{2} - \frac{1-\sqrt{z}}{2}\log_2\frac{1-\sqrt{z}}{2}.$$

Since the eq. (2) is a pure state, and the quantum discord reduces to entanglement entropy (*E*) in such case. So eq. (13) can also be expressed as $QD(\rho) = E(\rho) = -\alpha^2 \log_2 \alpha^2 - \beta^2 \log_2 \beta^2$. We find that eq. (13) is identical with eq. (14), therefore *QD*, *MID* and entanglement entropy are equal in this case. The characteristic of *QD* & *MID*, the first derivative of *QD* & *MID*, and the scaling behavior of

$\ln|dQD\&MID/dg|_{\min}$ have been displayed in Fig. 2. The *QD & MID* also can be used to discover the critical point $g_c$ correlated with QPTs through enough steps of QRG. The difference is that the saturated value of *QD & MID* is 1 for $0 \leq g < 1$. In addition, the singular behavior of *QD & MID* is more pronounced than negativity at the thermodynamic limit from Fig. 2(b). The scaling behavior between the minimum value of $dQD\&MID/dg$ and the size of the system also can be found in Fig. 2(c). The critical exponent $\theta$ of *QD & MID* are equal to negativity, i.e. $|dQD\&MID/dg|_{\min} \sim N$ with $\theta = 1$. We will check it further by measurement-induced nonlocality and geometric discord.

**Measurement-induced nonlocality and geometric discord.** Luo *et al*. introduced the measurement-induced nonlocality (*MIN*) in 2011 to quantify the quantum correlation[24, 29]

$$MIN_A(\rho_{AB}) := \max_{\Pi^A} \|\rho_{AB} - \Pi^A(\rho_{AB})\|^2, \tag{15}$$

here $\Pi^A(\rho)$ is the von Neumann measurements and $\rho^A = \sum_k \Pi_k^A \rho^A \Pi_k^A$.

Dakie *et al*. introduced the geometric measure of quantum discord as[22, 23, 29]

$$GQD(\rho) := \min_{\chi \in \Omega} \|\rho - \chi\|^2, \tag{16}$$

where $\Omega$ means the set of zero-discord states and $\|\rho - \chi\|^2 = tr(\rho - \chi)^2$ is the square of the Hilbert-Schmidt norm.

Applying these formulas to the eq. (5), one gets

$$MIN_A(\rho) = \begin{cases} (t_1^2 + t_2^2 + t_3^2 - \kappa_{\min})/4 & \text{if } x \equiv 0 \\ (t_1^2 + t_2^2)/4 & \text{if } x \neq 0 \end{cases}. \tag{17}$$
$$= 1/2(1+g^2),$$

$$GQD(\rho) = \left[t_1^2 + t_2^2 + t_3^2 + x^2 - \max(t_1^2, t_2^2, t_3^2, x^2)\right]/4 = 1/2(1+g^2), \tag{18}$$

where $x = \beta^2 - \alpha^2$, $t_1 = 2\alpha\beta$, $t_2 = -2\alpha\beta$, $t_3 = \beta^2 + \alpha^2$, $\kappa_{\min} = \min(t_1^2, t_2^2, t_3^2)$.

The results of *MIN&GQD*, the first derivative of *MIN&GQD*, and the scaling behavior of $\ln|dMIN\&GQD/dg|_{\min}$ are given in Fig. 3. The *MIN&GQD* also can detect the critical point. The singular behavior at the vicinity of the critical point can be seen in the Fig. 3(b) and the scaling behavior of $\ln|dMIN\&GQD/dg|_{\min}$ exist too. The critical exponent $\theta$ for $|dMIN\&GQD/dg|_{\min} \sim N^\theta$ is 1 and have no change.

**Quantum deficit.** For any bipartite state, the quantum deficit (*QDe*) is defined as the relative entropy of the state $\rho_{AB}$ with respect to its classically decohered counterpart $\rho_{AB}^d$ as below[28]

$$QDe_{AB} = S\left(\rho_{AB} \| \rho_{AB}^d\right) = Tr(\rho_{AB} \ln \rho_{AB}) - Tr\left(\rho_{AB} \ln \rho_{AB}^d\right). \tag{19}$$

The quantum deficit $QDe_{AB}$ determines the quantum excess of correlations in the state $\rho_{AB}$, with reference to its classical counterpart $\rho_{AB}^d$. The classical state $\rho_{AB}^d$ has the same marginal states $\rho_A$, $\rho_B$ as that of $\rho_{AB}$. It is diagonal in the eigenbasis $\{|a\rangle, |b\rangle\}$ of $\rho_A$, $\rho_B$ and the expression is

$$\rho_{AB}^d = \sum_{a,b} P_{ab} |a\rangle\langle a| \otimes |b\rangle\langle b|, \tag{20}$$

where $P_{ab} = \langle a,b|\rho_{AB}|a,b\rangle$ stands for the diagonal terms of $\rho_{AB}$ and $\sum_{a,b} P_{ab} = 1$.

So, it is easy to see that $Tr\left(\rho_{AB} \ln \rho_{AB}^d\right) = \sum_{a,b} P_{ab} \ln P_{ab}$, which leads to

$$QDe_{AB} = Tr(\rho_{AB} \ln \rho_{AB}) - Tr\left(\rho_{AB} \ln \rho_{AB}^d\right) = \sum_i \lambda_i \ln \lambda_i - \sum_{a,b} P_{ab} \ln P_{ab}. \tag{21}$$

where $\lambda_i$ signify the eigenvalues of the state $\rho_{AB}$. After some algebra, we can get $QDe$ for this model

$$QDe(\rho) = (\alpha^2 + \beta^2)\ln(\alpha^2 + \beta^2) - \alpha^2 \ln(\alpha^2) - \beta^2 \ln(\beta^2). \tag{22}$$

The performances of $QDe$, the first derivative of $QDe$, and the scaling behavior of $\ln|dQDe/dg|_{\min}$ are plotted in Fig. 4. Just like before, all curves in Fig. 4(a) cross each other at the critical point $g_c = 1$. The saturated values $QDe = 0.6931$ for $0 \leq g < 1$ and 0 for $g > 1$ after enough steps of renormalization. The singular behavior at the critical point and the scaling behavior of $\ln|dQDe/dg|_{\min}$ also can be observed if we use $QDe$ to quantify the quantum correlation. The exponent $\theta$ of $\ln|dQDe/dg|_{\min} \sim N^\theta$ still is 1. It is found that the critical exponent $\theta$ does not change with the variation of different measures.

**Bell violation.** The Bell violation can be adopted to prove the existence of quantum nonlocality. The following expression is the Bell operator corresponding to Clauser-Horne-Shimony-Holt (CHSH) inequality[29, 40-43]

$$B_{\text{CHSH}} = \boldsymbol{a} \cdot \boldsymbol{\sigma} \otimes (\boldsymbol{b} + \boldsymbol{b}') \cdot \boldsymbol{\sigma} + \boldsymbol{a}' \cdot \boldsymbol{\sigma} \otimes (\boldsymbol{b} - \boldsymbol{b}') \cdot \boldsymbol{\sigma}, \tag{23}$$

where $\boldsymbol{a}$, $\boldsymbol{a}'$, $\boldsymbol{b}$, $\boldsymbol{b}'$ are the unit vectors in $\mathbb{R}^3$, and $\boldsymbol{\sigma} = (\sigma_x, \sigma_y, \sigma_z)$. The CHSH inequality can be expressed as

$$B = |\langle B_{\text{CHSH}}\rangle_\rho| = |\text{Tr}(\rho B_{\text{CHSH}})| \leq 2. \tag{24}$$

The maximum violation of CHSH inequality is defined by

$$B_{\text{CHSH}}^{\max} = \max_{a,a',b,b'} \text{Tr}(\rho B_{\text{CHSH}}) \quad (25)$$

So, we can get the analytical result for this model as

$$B_{\text{CHSH}}^{\max} = 2\sqrt{(t_1^2 + t_2^2 + t_3^2 - \kappa_{\min})} = \frac{2\sqrt{2}}{\sqrt{g^2+1}}, \quad (26)$$

where the parameters are exactly the same as eq. (17).

The features of $B_{\text{CHSH}}$, the first derivative of $B_{\text{CHSH}}$, and the scaling behavior of $\ln|dB_{\text{CHSH}}/dg|_{\min}$ are shown in Fig. 5. In Fig. 5(a), it is found that the block-block correlations will violate the CHSH inequality and also exhibit QPTs at the critical point. The saturated values are different from before: one is $B_{CHSH} = 2.828$ for $0 \leq g < 1$ and the other is 2 for $g > 1$. The scaling behaviors of $\ln|dB_{\text{CHSH}}/dg|_{\min}$ convince us that the Bell violation also catches the critical behavior of the ITF due to the nonanalytic behavior of the Bell nonlocality[44]. The exponents for this property are $|dB_{\text{CHSH}}/dg|_{\min} \sim N$. The values of the critical exponents are identical with before.

## Discussions

In this study, we have combined the methods of quantum correlation and QRG theory to analyze the critical behavior of ITF model. Our results indicate that the critical behavior of the system can be described by quantum correlation or Bell violation. These quantum-information theoretic measures share the same singularity and the same finite-size scaling. The critical exponent which relates with the correlation length exponent will remain the value 1 even with the variation of different quantum correlation measures. Based on numerical computation, we have conjectured that the correlation length can be easily gotten by applying the QRG theory. The similarities and differences between each quantum correlation measures also are given. Furthermore, by applying QRG method the block-block correlations in ITF will violate the CHSH inequality.

**Acknowledgments**

This work was supported by the National Natural Science Foundation of China (Grant Nos. 11535004, 11375086, 1175085, 11120101005 and 11235001), by the 973 National Major State Basic Research and Development of China grant No. 2013CB834400, and Technology Development Fund of Macau grant No. 068/2011/A., by the Project Funded by the Priority Academic Program Development of Jiangsu Higher Education Institutions (PAPD), by the Research and Innovation Project for College Postgraduate of JiangSu Province (Grants No. KYZZ15_0027).

**Author Contributions**

M.Q. carried out the calculations and plotted the figures. M.Q., Z.Z.R. and X.Z. discussed the results and wrote the paper. All authors reviewed the manuscript.

**Additional Information**

**Competing financial interests:** The authors declare no competing financial interests.

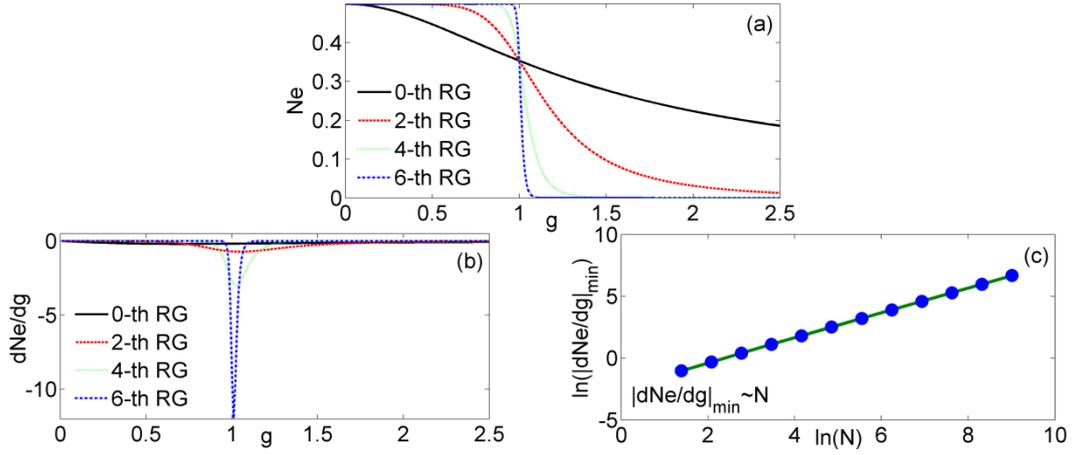

**Figure 1.** The negativity (a) and the first derivative of negativity (b) of the model versus *g* at different quantum renormalization group steps. The logarithm of the absolute value of minimum $|dNe/dg|_{min}$ (c) in terms of system size $\ln(N)$.

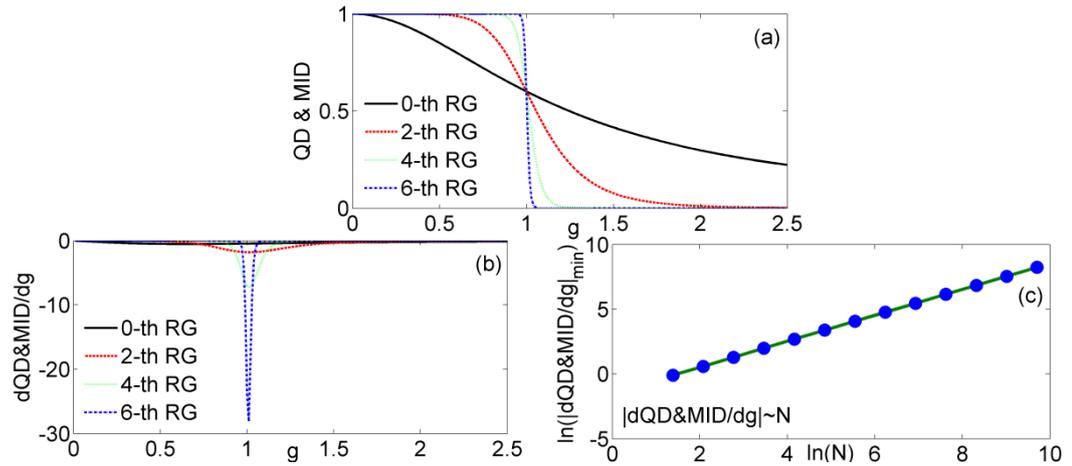

**Figure 2.** The *QD & MID* (a) and the first derivative of *QD & MID* (b) versus *g* at different quantum renormalization group steps. The logarithm of the absolute value of minimum $\ln|dQD\&MID/dg|_{min}$ (c) in terms of system size $\ln(N)$.

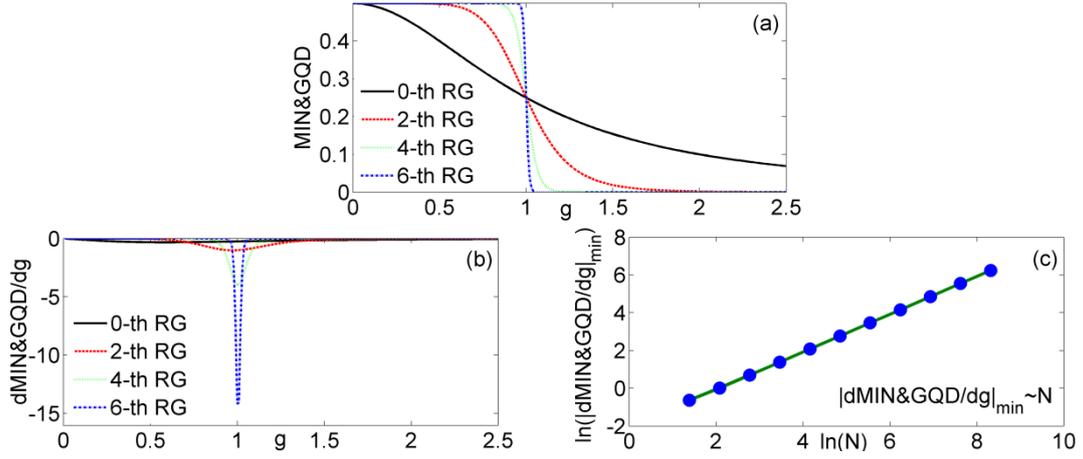

**Figure 3.** The $MIN\&GQD$ (a) and the first derivative of $MIN\&GQD$ (b) of the model versus $g$ at different quantum renormalization group steps. The logarithm of the absolute value of minimum $\ln|dMIN\&GQD/dg|_{\min}$ (c) in terms of system size $\ln(N)$.

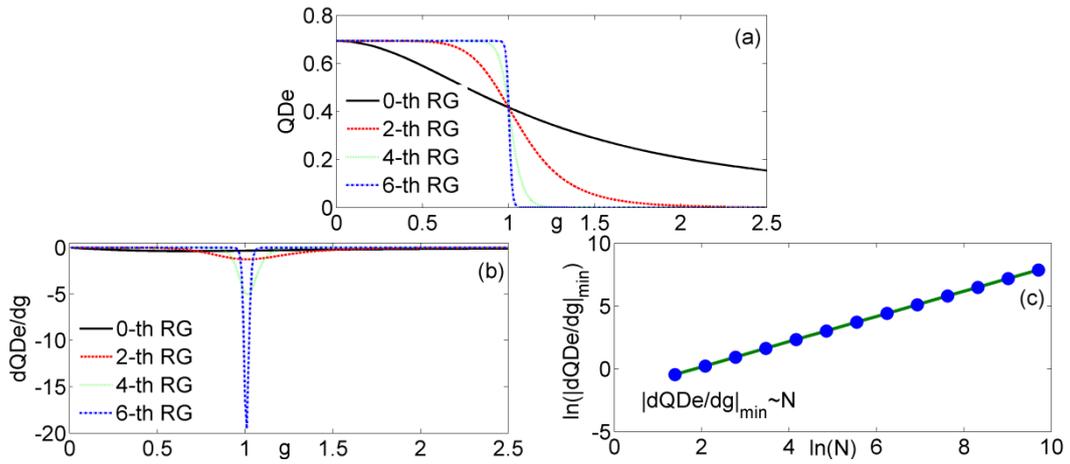

**Figure 4.** The $QDe$ (a) and the first derivative of $QDe$ (b) of the model versus $g$ at different quantum renormalization group steps. The logarithm of the absolute value of minimum $\ln|dQDe/dg|_{\min}$ (c) in terms of system size $\ln(N)$.

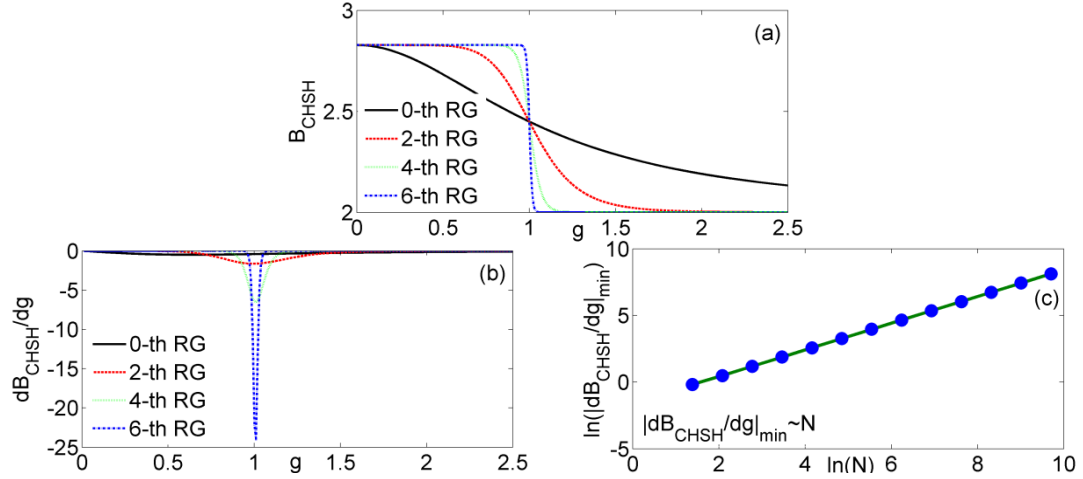

**Figure 5.** The $B_{CHSH}$ (a) and the first derivative of $B_{CHSH}$ (b) of the model versus $g$ at different quantum renormalization group steps. The logarithm of the absolute value of minimum $\ln|dB_{CHSH}/dg|_{min}$ (c) in terms of system size $\ln(N)$.